\documentclass[12pt]{article}
\usepackage[margin=1in]{geometry}
\usepackage{setspace}
\usepackage{natbib}
\usepackage{newfloat}
\usepackage{amsmath}
\usepackage{amsfonts}
\usepackage{amssymb}
\usepackage{lineno}
\usepackage[utf8]{inputenc}
\usepackage{physics}
\usepackage{bm}
\newcommand{\pdev}[2]{\frac{\partial #1}{\partial #2}}
\newcommand{\sqmatrix}[1]{\begin{bmatrix}#1\end{bmatrix}}

\DeclareFloatingEnvironment[name={Extended Data Figure}]{suppfigure}
\newcommand{\mathsfbi}[1]{\boldsymbol{\mathsf{#1}}}
\usepackage[english]{babel}

\usepackage{graphicx}
\usepackage{xcolor}

\definecolor{darkgreen}{rgb}{0.0, 0.5, 0.0}
\date{}

\begin{document}

\title{Thermocapillary stabilization of Liquid Space Telescopes}

\author{
Ryan Engle$^{1}$ and Valeri Frumkin$^{1}$\\
$^{1}$Department of Mechanical Engineering, Boston University, Boston, MA, USA
}

\maketitle

\begin{abstract}
We investigate how spatially uniform, time-modulated heating can be used to stabilize a thin liquid mirror in microgravity. Periodic heating generates oscillatory thermocapillary flows that suppress the instability associated with sustained heating, allowing the free surface to relax through the combined effects of thermal radiation, thermocapillary transport, and capillarity. Using multiple-scale analysis of a long-wave model, we derive the coupled surface-temperature evolution over many heating cycles. A subsequent linear stability analysis yields the stability criteria and decay times of individual surface modes. Numerical simulations validate the analytical predictions and demonstrate stabilization of meter-scale surface deformations over timescales of days.
\end{abstract}
\begin{figure}[h!]
    \centering
    \includegraphics[width=0.5\linewidth]{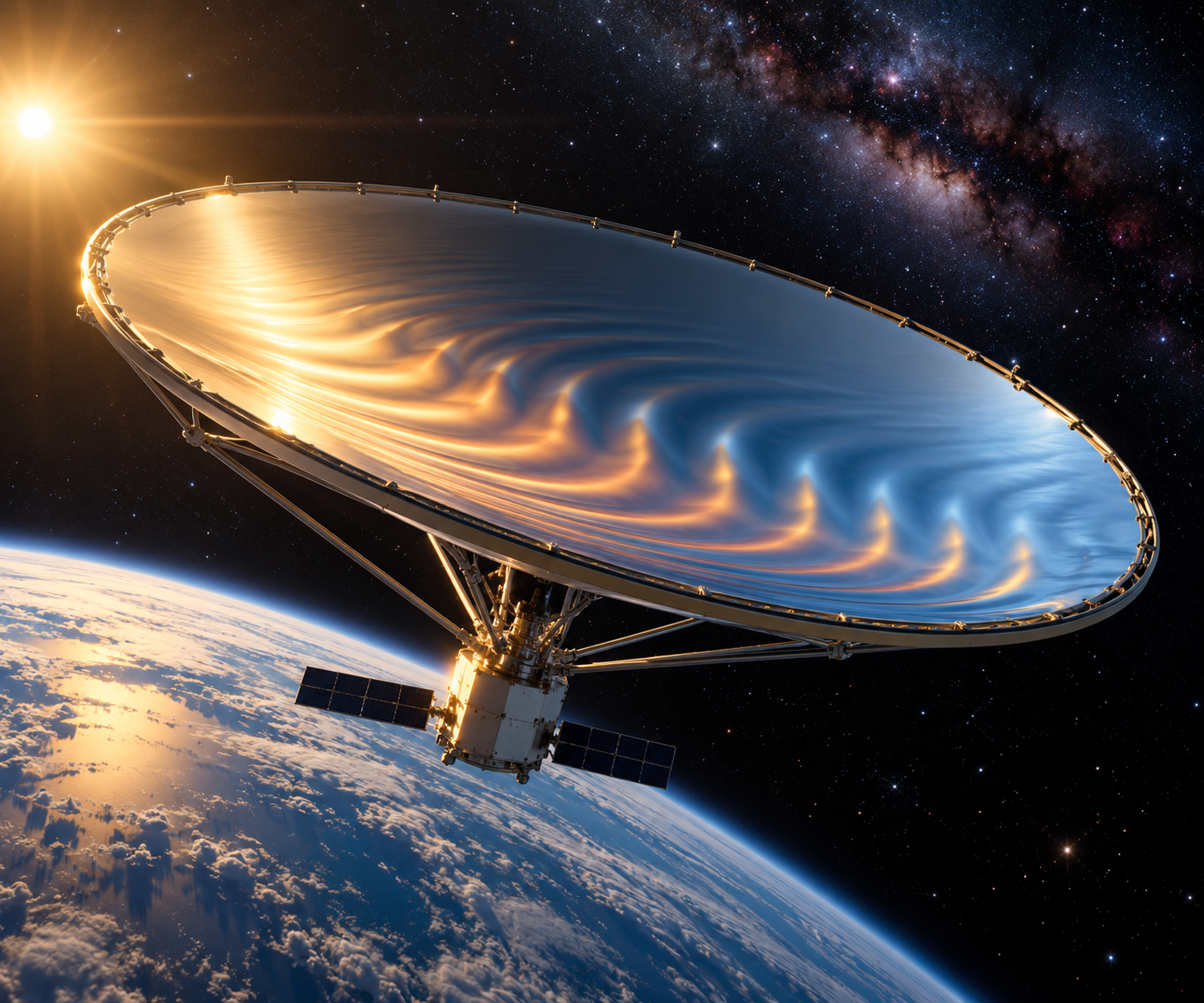}
\end{figure}
\newpage
\section{Introduction}

Current demands for direct exoplanet observation require apertures tens of meters in diameter to differentiate the planet from its host star \citep{seager_future_2014, gong_optical_2019}. The James Webb Space Telescope (JWST), launched in 2021, is currently the largest operational space observatory with a 6.6-m primary mirror \citep{gardner_james_2023}. To fit inside the Ariane 5 rocket, the primary mirror was segmented into 18 hexagonal units that folded together during launch. Although segmentation enables easier transport, precise alignment of the individual mirror segments is highly challenging at this scale. The JWST uses state-of-the-art 7-axis actuators with nanometer-scale resolution over millimeter-scale range for mirror alignment  \citep{mcelwain_james_2023}. While mirror diameters up to 8-m were considered, the complexity of implementing an effective alignment system limited the JWST to the final 6.6-m design  \citep{gardner_james_2023}. Therefore, new technologies are needed to reach the tens-of-meters scale required for direct exoplanet observation.

Spinning Liquid Mirror Telescopes (SLMTs) offer an alternative to solid segmented apertures by using a self-shaping reflective surface. First proposed by Newton and constructed in the 19th century, SLMTs form a mirror by rotating a reflective liquid around the axis of gravity. The balance between centrifugal forces and gravity shapes the liquid into a parabolic mirror without the challenges associated with the fabrication and polishing of a solid optical surface \citep{borra_liquid_1997}. However, SLMTs have limited applicability due to their fixed orientation parallel to gravity. The Zenith project \citep{nayak_overview_2024} aims to circumvent this by stabilizing an SLMT through electromagnetic actuation, as was first suggested by \citet{ragazzoni_liquid_1994} and \citet{shuter_wide_1994}. However, precise field control across the full diameter requires current magnitudes and coil precision beyond current technology \citep{comstock_feasibility_2025}. As a result, groups participating in the Zenith project have pursued tileable actuation units to control individual mirror sections, reducing the required current magnitude and coil precision. However, this approach introduces surface errors associated with the size of the unit \citep{rowlands_development_2024} and would require hundreds or thousands of units across a full-sized mirror. Consequently, substantial development is still needed to realize scalable electromagnetic SLMTs.

The Fluidic Telescope (FLUTE) project, which this work is associated with, seeks to develop a large liquid mirror observatory in space. The proposed primary mirror will be formed from a reflective liquid layer a few millimeters thick, coating a solid surface tens-of-meters in diameter. In the absence of gravity, the dynamics of the thin layer are dominated by capillary pressure, shaping the surface into a spherical minimum energy state \citep{frumkin_fluidic_2021, luria_-space_2026}. This allows the mirror to form independent of orientation and without the need for active rotation. However, recent studies have shown that the time of relaxation scales unfavorably with mirror size due to capillary pressure's dependence on local mean curvature, making passive equilibration prohibitively slow for long-wave deformations  \citep{dietrich_equilibrium_2026, gabay_fluid_2026}. For this reason, an additional mechanism is needed to stabilize long-wave deformations that emerge from initial deployment or telescope maneuvers such as slewing.

Thermocapillary flow is a promising mechanism for accelerating the free surface stabilization process.  \citet{pearson_convection_1958} first studied this mechanism in the context of thin films, who described the onset of thermocapillary instability due to temperature-driven surface tension gradients.  
\citet{scriven_cellular_1964} extended this works to a more realistic setting with convective cooling, finite surface tension, and viscous effects. The current long-wave model, first derived by \citet{davis_thermocapillary_1987}, uses a lubrication approximating to construct a surface evolution equation, and has since been used to study thermocapillary flow in combination with other thin-film dynamics \citep{oron_long-scale_1997}.  Since surface temperature typically increases with decreasing film thickness \citep{oron_long-scale_1997}, thermocapillary flow act as a destabilizing mechanism, leading to film rupture unless saturated by other nonlinear effects. Careful control of this balance allows thermocapillary flow to used to create predictable pattern formation \citep{mcleod_experimental_2011, nazaripoor_thermally_2018} and to enable active surface control \citep{eshel_programmable_2022} in thin-film systems. Prescribed temperature is usually assumed at the boundary between the liquid film and the underlying substrate, due to the slow evolution of surface deformations when compared to the much faster thermal conduction time scales. While this simplifying assumption applies for many thin-film systems, it does not hold for FLUTE. The liquid layer is significantly thicker at the millimeter length scale, greatly increasing the thermal conduction time, and the applied heating is planned to be temporally modulated. These two factors prevent the system temperature from settling to a set profile, making a fixed temperature boundary condition ill-suited for the FLUTE system. Therefore, the thermal boundary condition at the substrate is best modeled as a modulated heat flux. 

\citet{or_effects_2002} were the first to study modulated heat flux and found stability criteria for very long wavelengths under an insulated free surface and sinusoidal heating. The modulation provided additional stabilization that could overcome the traditional thermocapillary destabilization even in the absence of gravity. \citet{mikishev_long-scale_2010} were able to extend this to all wavelengths using multiple scale analysis to separate the thermal evolution and surface evolution timescales. However, the insulated free surface and net-zero sinusoidal heat flux are highly restrictive and do not apply to real physical systems.

In this work, we demonstrate the stabilization of a large-scale liquid mirror in microgravity using modulated thermocapillary flow. We first derive an axisymmetric long-wave model for a spherical thin layer subject to surface radiation and modulated by uniform periodic heat pulses. Multiple scale analysis is used to derive the long-time dynamics of the free surface and temperature, extending the method presented in \citet{mikishev_long-scale_2010} to non-sinusoidal modulation and boundary conditions that weakly depend on temperature. The subsequent linear stability analysis extracts the oscillation and decay times of individual surface modes based on system parameters, as well as the stability criteria for the system. The resulting solution is then compared to a finite difference simulation of the long-wave model for different perturbation modes. 

\section{Theory}

\begin{figure}[h]
    \centering
    \includegraphics[width=\linewidth]{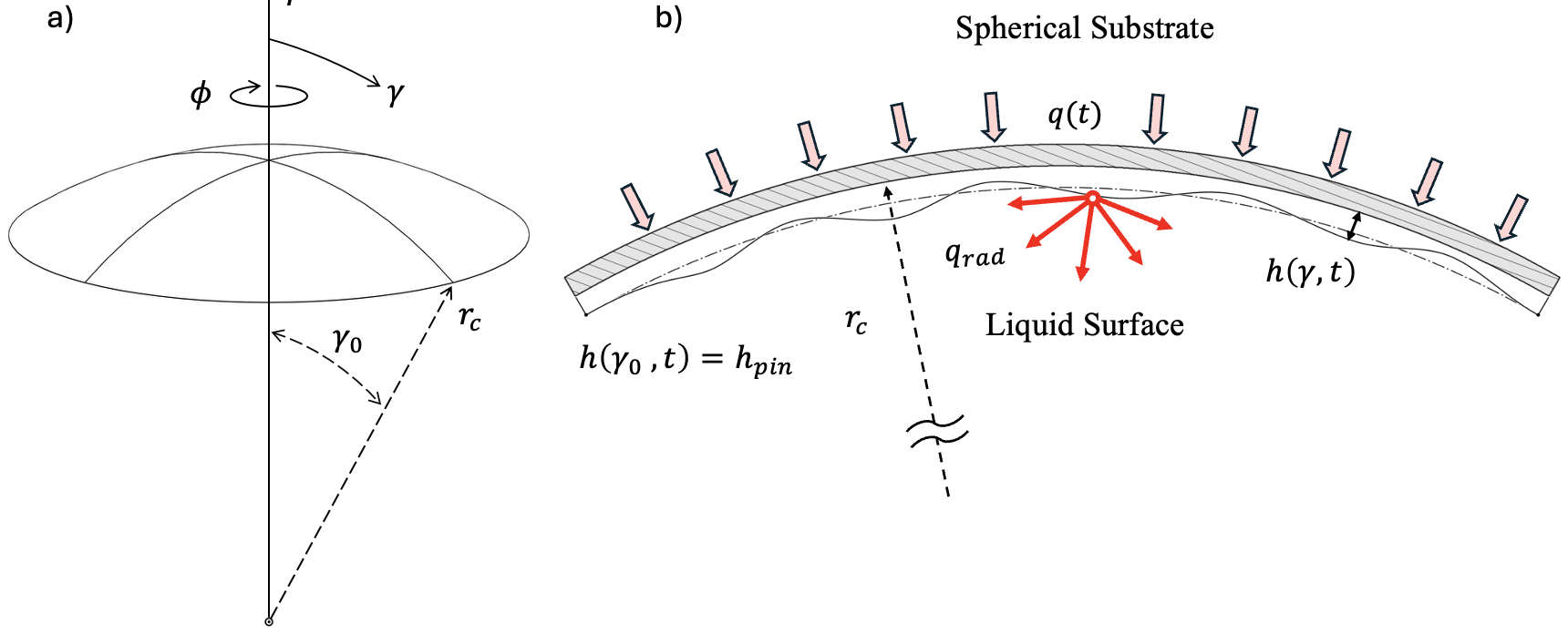}
    \caption{Schematic illustration of the large liquid mirror. (a) A 3D view of the mirror with spherical coordinates. (b) Cross-section view of the shell-liquid system. Uniform heating is applied at the substrate and the free surface emits radiation.}
    \label{fig:Mirror Schematic}
\end{figure}

We consider a system consisting of a thin liquid layer coating the inside of a spherical frame with radius of curvature $r_c$ spanning a polar angle of $\pm\gamma_0$. The system is assumed to be axisymmetric and is described by spherical coordinates with a radial distance $r$, a polar angle $\gamma$, and an azimuthal angle $\phi$ (Figure \ref{fig:Mirror Schematic}). The thickness of the layer $h(\gamma, t)$ is pinned at the outer edge to a thickness of $h_{pin}$, with fluid  able to flow freely past the boundaries in and out of a reservoir (see Supplementary Information for discussion of polar boundary conditions). The free surface radiates energy to both the background of space as well as other parts of the liquid mirror. Finally, a spatially uniform time-modulated heat flux $q(t)$ is applied at the solid-liquid interface, yielding thermocapillary stresses on the free surface of the liquid layer.

\subsection{Governing equations}

The fluid dynamics within the liquid mirror are described by the incompressible Navier-Stokes equations and the thermodynamics by the advection-diffusion heat equation:

\begin{gather}
    \pdev{\boldsymbol{u}}{t} + (\boldsymbol{u\cdot \nabla}) \boldsymbol{u} = -\frac{1}{\rho} \boldsymbol{\nabla} p + \nu \nabla^2 \boldsymbol{u} \label{eq:NS Dim}, \\
    \boldsymbol{\nabla \cdot u} = 0 \label{eq:Continuity Dim}, \\\pdev{T}{t}  + (\boldsymbol{u\cdot\nabla})T = \alpha\nabla^2T \label{eq:Heat}.
\end{gather}

The free surface of the mirror, located at $r_s(\gamma, t) = r_c - h(\gamma, t)$, moves according to the kinematic boundary condition where the surface velocity matches the normal fluid velocity:

\begin{gather}
    \boldsymbol{u} \boldsymbol{\cdot \hat n} \Big|_{r = r_s(\gamma, t)} = -(\boldsymbol{\hat r \cdot \hat n})\pdev{h}{t}. \label{eq:Kinematic Dim}
\end{gather}

At the free surface, a stress balance is applied to match the normal and tangent stresses at the liquid surface to capillary pressure and surface tension gradients, respectively:

\begin{gather}
    \mathsfbi{\bar T} \boldsymbol{\cdot \hat n} \Big|_{r = r_s(\gamma, t)} = -\boldsymbol{\hat n}\cdot  \sigma (\boldsymbol{\nabla \cdot \hat n}) + \boldsymbol{\nabla_s} \sigma \label{eq:Stress Dim},
\end{gather}

where $\boldsymbol{\nabla_s} \equiv \boldsymbol{\nabla} - \boldsymbol{\hat n} (\boldsymbol{\hat n}  \cdot \boldsymbol{\nabla})$ is the gradient parallel the free surface. The surface tension gradients in Equation \ref{eq:Stress Dim} arise from temperature gradients due to the thermocapillary effect. The relation between temperature and surface tension is estimated as linear around the average free surface temperature $T_{ref}$:

\begin{gather}
    \sigma = \sigma_{ref} - \beta(T-T_{ref}) \Big|_{r = r_s(\gamma, t)}, \label{eq:Thermocapillary Effect}
\end{gather}

where $\sigma_{ref}$ is the surface tension at $T=T_{ref}$ and $\beta \equiv -d\sigma/dT>0$ is the thermocapillary coefficient. The free surface exchanges energy via radiation with both the background of space as well as other sections of the mirror surface. The background of space is assumed to be a black body with a temperature $T_{bg}$. The net energy transfer at a point on the free surface, $r_s(\gamma,t)$, can be determined by integrating the net energy flux to and from the surface $A'$ containing all points visible to $r_s(\gamma,t)$:

\begin{gather}
    -k\boldsymbol{\nabla}T \boldsymbol{\cdot \hat n} \Big|_{r = r_s(\gamma, t)} = \left[\varepsilon\sigma_{sb} T^4 -\iint_{A'} \frac{\boldsymbol{(s \cdot \hat n)(s\cdot \hat n')}}{\pi\lVert\boldsymbol{s}\rVert^4} a\varepsilon' \sigma_{sb} T'^4 dA'  \right]_{r = r_s(\gamma, t)}, \label{eq:Radiation} 
\end{gather}

where $\varepsilon$ is the emissivity of the surface (assumed to be constant for the mirror), $a$ is the proportion of emitted light from $dA'$ that is absorbed (also assumed to be constant), $\sigma_{sb}$ is the Stefan-Boltzmann constant, and $\boldsymbol{s}$ is the distance vector between $r(\gamma, \phi)$ and the surface element $dA'$ with temperatures $T$ and $T'$ respectively. 

At the solid-liquid interface, $r=r_c$, we assume a no-slip condition on the fluid velocity:

\begin{gather}
    \boldsymbol{u}\Big|_{r=r_c} = (0, 0). \label{eq:No Slip}
\end{gather}

Finally, we apply a spatially uniform modulated heat flux at the solid-liquid interface. The heating alternates between an on-period, with a magnitude $q_0$ lasting a period of $t_h$, and an off-period where the boundary is assumed to be insulating, for a total cycle time of $t_{ch}$. During these off-periods, radiation cools the mirror surface with thinner regions cooling faster than thicker regions. As will be shown, this generates stabilizing oscillations as fluid is drawn to the thinner regions, significantly reducing the thermocapillary instability that arises from heating. The heat flux boundary condition at the solid-liquid interface is thus represented as an infinite series of step functions:

\begin{gather}
    -k\boldsymbol{\nabla}T \boldsymbol{\cdot \hat r} \Big|_{r=r_c} = -q_0\sum_{n=0}^{\infty}(\Theta (t-nt_{ch}) - \Theta(t-t_h - nt_{ch})). \label{eq:Heat Cycle}
\end{gather}

\subsection{Long-wave model}

We non-dimensionalize the governing equations using the characteristic transverse distance $w$, the characteristic radial distance $h_{pin}$, the reference temperature $T_{ref}$, and the polar span $\gamma_0$. We define $\epsilon$ as the ratio between the radial and transverse length scales, $\zeta$ as the characteristic temperature gradient from balancing radiative and conductive heat transfer, and $u_t$ as the characteristic thermocapillary velocity from that gradient:

\begin{gather}
    w \equiv r_c \gamma_0, \quad \epsilon \equiv \frac{h_{pin}}{w} \sim 10^{-3}, 
    \quad 
    \zeta \equiv \epsilon \frac{\varepsilon \sigma_{sb}T_{ref}^4}{k},
    \quad
    u_t \equiv \epsilon\frac{\beta \zeta w}{\mu}.
\end{gather}

We use these characteristic values to introduce the following non-dimensional variables:

\begin{gather}
    t = \frac{h_{pin}^2}{\alpha} \tau, \quad r = r_c + h_0R, \quad \gamma = \gamma_0 \Gamma, \quad h = h_{pin}H, \\
    u_r = \epsilon u_t \bar U_r, \quad u_\gamma = u_t \bar U_\gamma, \\
    p = \frac{\mu u_t}{\epsilon h_{pin}}P, \quad \sigma = \sigma_{ref} + \Sigma \cdot \beta \zeta w, \quad \theta = \frac{T - T_{ref}}{\zeta w}.
\end{gather}

From this process, we obtain an additional set of dimensionless parameters that describe the overall dynamics of the system:

\begin{gather}
    \bar P \equiv \frac{\alpha}{\epsilon \nu} \sim 1, \quad \bar S \equiv \epsilon^3 \frac{\sigma_{ref}}{\mu u_t} \sim 10^{-3}, \\ \bar M \equiv \epsilon^2 \frac{h_{pin} \beta \zeta w}{\mu \alpha} \sim 10^{-5}, \quad \xi \equiv \frac{h_{pin}\varepsilon \sigma_{sb}T_{ref}^3}{k} \sim 10^{-3}, \quad \upsilon \equiv \frac{T_{bg}^4}{\varepsilon T_{ref}^4} \sim 10^{-4},
\end{gather}

where $\bar P$ is the scaled inverse Prandtl number, $\bar S$ is the scaled inverse capillary number representing the ratio between capillary pressure and the thermocapillary effect, $\bar M$ is the scaled Marangoni number representing the ratio between the time scales of radial thermal conduction and thermocapillary mass transfer, $\xi$ is the radiation parameter representing the ratio between the thermal conduction time scale and the settling time for the surface temperature through radiation, and $\upsilon$ is the ratio between the background radiation and the mirror surface radiation.

To obtain a long-wave evolution equation, we proceed by solving for the temperature and surface evolution at leading order in $\epsilon$. Transverse conduction is eliminated from the advection-diffusion equation \ref{eq:Heat}, yielding:

\begin{gather}
    \pdev{\theta}{\tau} + \bar M \left(\bar U_r \pdev{\theta}{R} + \bar U_\gamma \pdev{\theta}{\Gamma}  \right) = \pdev{^2\theta}{R^2}, \label{eq: ND Temperature Evolution}
\end{gather}

and the applied heat flux \ref{eq:Heat Cycle} is given by:

\begin{gather}
    \pdev{\theta}{R}\Big|_{R=0}  = Q_0\sum_{n=0}^\infty \left(\Theta(\tau-n\tau_{ch}) - \Theta(\tau - \tau_h - n\tau_{ch}) \right), \label{eq:ND Heat Cycle} \\
    Q_0 \equiv \frac{q_0}{\varepsilon \sigma_{sb} T_{ref}^4} \sim 30,\quad \tau_h \equiv \frac{\alpha t_h}{h_{pin}^2} \sim 0.1, \qquad \tau_{ch} \equiv \frac{\alpha t_{ch}}{h_{pin}^2} \sim 10,
\end{gather}

where $Q_0$, $\tau_h$, and $\tau_{ch}$ are the scaled magnitude, heating time, and cycle time respectively. As the heating changes at a time scale similar to radial conduction, the temperature profile will not settle and instead oscillate as the heating turns on and off. For this reason, we study the system at the radial conduction time scale where these oscillations can be modeled.

The radiative boundary condition \ref{eq:Radiation} can be greatly simplified by assuming the mirror forms a convex surface. Under this assumption, which holds for small disturbances at long wavelengths, each point on the mirror surface can see both the full background and all other points on the mirror. The view factor can then be reduced at the leading order in $\epsilon$ to the view factor inside a spherical shell (see Supplementary Information for more details):

\begin{gather}
    \frac{\boldsymbol{(s \cdot \hat n)(s\cdot \hat n')}}{\pi\lVert\boldsymbol{s}\rVert^4} = \frac{1}{4\pi r_0^2} + O(\epsilon).
\end{gather}

This removes any spatial component from the view factor, simplifying the radiative boundary condition to the balance between the emitted energy at each point and the absorbed radiation expressed as a constant heat flux:

\begin{gather}
    \pdev{\theta}{R}\Big|_{R=-H} =\left[(1+\xi \theta)^4 -  \frac{a \gamma_0}{2}\int_0^{1} (1+\xi \theta)^4\sin \gamma_0\Gamma \, d\Gamma - a\upsilon \left(\frac{1+\cos \gamma_0}{2} \right)\right]_{R=-H}.
\end{gather}

As $\xi \ll 1$, the free surface temperature will slowly settle over the course of many heating cycles until the average radiation emitted matches the average heating applied. We refer to this process as radiative settling.

At the leading order in $\epsilon$, the dimensionless velocity profile is set by the capillary pressure, which grows stronger for smaller wavelengths, and the thermocapillary effect at the fluid surface:

\begin{gather}
    \bar U_r = \bar \nabla_2 \cdot \left(\bar S \left(\frac{1}{6}R^3 + \frac{1}{2} H R^2 \right) \left(2 \gamma_0^2 H + \bar \nabla_2^2 H\right) - \frac{1}{2}R^2 \bar \nabla_2 \theta \Big|_{R=-H}\right),
    \label{eq:LW Radial Velocity}
    \\
    \bar U_\gamma = - \bar S \left(\frac{1}{2}R^2 + H R \right) \left(2 \gamma_0^2H + \bar \nabla_2^2 H\right) + R \bar \nabla_2 \theta \Big|_{R=-H},
    \label{eq:LW Polar Velocity}
    \\
    \bar \nabla_2 f\equiv \pdev{f}{\Gamma}, \qquad \bar \nabla_2 \cdot f\equiv \frac{1}{\sin \gamma_0 \Gamma} \pdev{}{\Gamma}\left( f\sin \gamma_0 \Gamma\right),
    \label{eq:Scaled Del Definition}
\end{gather}

where $\bar \nabla_2$ is the del operator on a unit sphere with the scaled polar angle $\Gamma$. The inherent spherical curvature of the substrate provides an additional contribution to the capillary pressure. As the film thickness increases, the negative pressure from the substrate curvature also increases and draws in fluid from thinner regions. 

Integrating the continuity equation \ref{eq:Continuity Dim} along the radial direction and implementing the kinematic boundary condition \ref{eq:Kinematic Dim}, we obtain a long-wave evolution equation for the liquid's free surface (see Supplementary Information for detailed derivation)

\begin{gather}
    \pdev{H}{\tau} +  \bar M \bar \nabla_2 \cdot \left(\frac{1}{3}\bar SH^3 \bar \nabla_2 (2 \gamma_0^2 H + \bar \nabla_2^2 H) - \frac{1}{2}H^2 \bar \nabla_2 \theta \Big|_{R=-H} \right) = 0.
\end{gather}

\subsection{Short-Term and Long-Term Evolution}

We separate the dynamics between the fast evolution of the heating cycles, and the long-term evolution of radiative settling and mass transfer over the course of many cycles using multiple scale analysis. We define the scaled inverse capillary number $S \equiv \bar S \gamma_0^2$ and scaled Marangoni number $M \equiv \bar M \gamma_0^2$, and obtain the transformed fluid velocities and radiation boundary condition:

\begin{gather}
    U_r = \nabla_2 \cdot \left(S \left(\frac{1}{6}R^3 + \frac{1}{2} H R^2 \right) \left(2H + \nabla_2^2 H\right) - \frac{1}{2}R^2 \nabla_2 \theta \Big|_{R=-H}\right),
    \\
    U_\gamma = - S \left(\frac{1}{2}R^2 + H R \right) \left(2H + \nabla_2^2 H\right) + R \nabla_2 \theta \Big|_{R=-H},
    \\
    \pdev{\theta}{R}\Big|_{R=-H} =\left[(1+\xi \theta)^4 - \frac{a}{2}\int_0^{\gamma_0} (1+\xi \theta)^4\sin \gamma \, d\gamma - \left(\frac{1+\cos \gamma_0}{2} \right)\upsilon\right]_{R=-H},
    \\
    \nabla_2 f\equiv \pdev{f}{\gamma}, \qquad \nabla_2 \cdot f\equiv \frac{1}{\sin \gamma} \pdev{}{\gamma}\left( f\sin \gamma\right).
\end{gather}

Next, we express the scaled Marangoni number in terms of the radiation time scale through the ratio $\kappa \equiv M / \xi \sim 10^{-3}$:

\begin{gather}
    \pdev{H}{\tau} +  \kappa \xi \nabla_2 \cdot \left(\frac{1}{3} SH^3  \nabla_2 (2H + \nabla_2^2 H) - \frac{1}{2}H^2 \nabla_2 \theta \Big|_{R=-H} \right) = 0, 
    \label{eq:LW Surface Normal Angle}
    \\
    \pdev{\theta}{\tau} + \kappa \xi \left(U_r \pdev{\theta}{R} + U_\gamma \pdev{\theta}{\gamma}  \right) = \pdev{^2\theta}{R^2}. \label{eq:LW Temperature Normal Angle}
\end{gather}

We now expand our surface, temperature, and time variables in terms of $\xi$:

\begin{gather}
    H = H_0 + \xi H_1 +O(\xi^2) \quad \theta=\theta_0 +\xi\theta_1 + O(\xi^2) \quad \pdev{}{\tau} = \pdev{}{\tau_0} + \xi\pdev{}{\tau_1} + O(\xi^2).
\end{gather}

At the leading order, we find that the surface $H_0$ is independent of $\tau_0$, the convection terms are eliminated, and the surface radiation reduces to a constant uniform heat flux $V$:

\begin{gather}
    \pdev{H_0}{\tau_0} = 0, \\
    \pdev{\theta_0}{\tau_0} = \pdev{^2 \theta_0}{R^2}, \\
    \pdev{\theta_0}{R} \Big|_{R=-H_0} = 1-a\left(\frac{1-\cos\gamma_0}{2} \right) - a\upsilon \left(\frac{1+\cos \gamma_0}{2} \right) \equiv V ,\\
    \pdev{\theta_0}{R}\Big|_{R=0}  = Q_0\sum_{n=0}^\infty \left(\Theta(\tau_0-n\tau_{ch}) - \Theta(\tau_0 - \tau_h - n\tau_{ch}) \right).
\end{gather}

To balance the applied heating with the average radiative heat flux at $T_{ref}$, we set $Q_0 =V \tau_{ch}/\tau_h$. This allows us to solve for the temperature profile $\theta_0 = \breve \theta_0 (R, \gamma, \tau_0) + \tilde \theta_{0} (\gamma, \tau_1)$ by separating the radial profile $\breve \theta_0(R, \gamma,  \tau_0)$ and the long-term transverse profile $\tilde \theta_0(\gamma, \tau_1)$. As $\tilde \theta_0$ is independent of $R$ and $\tau_0$, it drops out at the leading order in $\xi$. The solution for $\breve \theta_0$ is found to be:

\begin{gather}
    \breve \theta_0 (\tau_0, R)= V\left(- \left[\frac{\tau_0 \mod{\tau_{ch}}}{H_0} + \frac{R^2}{2H_0} - \frac{H_0}{6} \right] \right. 
    \nonumber \\
    + 2H_0 \sum_{m=1}^\infty \frac{(-1)^m}{m^2\pi^2} \cos\left(\frac{m\pi}{H_0} R \right)\exp\left(-\frac{m^2\pi^2}{H_0^2} \tau_0  \right) 
    \nonumber\\ 
    + 2 \frac{\tau_{ch}}{\tau_h} \sum_{m=1}^\infty \frac{H_0}{m^2\pi^2} \cos\left(\frac{m\pi}{H_0} R \right) \left( \frac{\exp\left(\frac{m^2\pi^2}{H_0^2} \tau_h  \right) - 1}{\exp\left(\frac{m^2\pi^2}{H_0^2} \tau_{ch}  \right) - 1}\right) 
    \nonumber \\
    \times \left(\exp\left(-\frac{m^2\pi^2}{H_0^2} \left( \tau_0 \mod{\tau_{ch}}\right) \right) - \exp\left(-\frac{m^2\pi^2}{H_0^2} \tau_0 \right) \right)
    \nonumber \\
    \left.+ \frac{\tau_{ch}}{\tau_h} \cdot \begin{cases}
        \frac{\tau_0 \mod{\tau_{ch}}}{H_0} + \frac{(R+H_0)^2}{2H_0} - \frac{H_0}{6} -2 H_0 \sum_{m=1}^\infty  \frac{1}{m^2\pi^2} \cos\left(\frac{m\pi}{H_0} R \right) \\ \times \exp\left(-\frac{m^2\pi^2}{H_0^2} \left(\tau_0 \mod{\tau_{ch}} \right) \right) & \tau_0 \mod{\tau_{ch}} \leq \tau_h \\
        \\
        \frac{\tau_h}{H_0} + 2H_0 \sum_{m=1}^\infty \frac{1}{m^2\pi^2}  \cos\left(\frac{m\pi}{H_0} R \right)  \left(\exp\left(\frac{m^2\pi^2}{H_0^2} \tau_h \right) - 1  \right) \\\times \exp\left(-\frac{m^2\pi^2}{H_0^2} \left(\tau_0 \mod{\tau_{ch}} \right) \right) & \tau_0 \mod{\tau_{ch}} \geq \tau_h
    \end{cases} \right).
\end{gather}

At $O(\xi)$,  we obtain the coupled long-term and short-term evolution for the free surface and temperature:

\begin{gather}
    \pdev{H_0}{\tau_1} + \pdev{H_1}{\tau_0} +  \kappa \nabla_2 \cdot \left(\frac{1}{3} SH_0^3  \nabla_2 (2H_0 + \nabla_2^2 H_0) - \frac{1}{2}H_0^2 \nabla_2 \theta_0 \Big|_{R=-H_0} \right) = 0, \label{eq:Mixed Surface Evolution}
    \\
    \pdev{\theta_0}{\tau_1} + \pdev{\theta_1}{\tau_0} + \kappa \left(U_{r0} \pdev{\theta_0}{R} + U_{\gamma 0} \pdev{\theta_0}{\gamma}  \right) = \pdev{^2\theta_1}{R^2}, \label{eq:Mixed Temperature Evolution}
    \\
    U_{r0} = \nabla_2 \cdot \left(S \left(\frac{1}{6}R^3 + \frac{1}{2} H_0 R^2 \right) \left(2H_0 + \nabla_2^2 H_0\right) - \frac{1}{2}R^2 \nabla_2 \theta_0 \Big|_{R=-H_0}\right), \label{eq:Mixed Radial Velocity}
    \\
    U_{\gamma0} = - S \left(\frac{1}{2}R^2 + H_0 R \right) \left(2H_0 + \nabla_2^2 H_0\right) + R \nabla_2 \theta \Big|_{R=-H_0}, \label{eq:Mixed Transverse Velocity}
    \\
    \left[\pdev{\theta_1}{R} - H_1 \pdev{^2 \theta_0}{R^2} \right]_{R=-H} =4\left[\theta_0 -  \frac{a}{2}\int_0^{\gamma_0} \theta_0 \sin \gamma \, d\gamma\right]_{R=-H_0}, \label{eq:Mixed Radiation}
    \\
    \pdev{\theta_1}{R}\Big|_{R=0} = 0. \label{eq:Mixed Solid-Liquid Interface}
\end{gather}

We then isolate the long-term dynamics of $H_0$ and $\tilde \theta_0$ through a matching procedure. First, we isolate the sustained surface oscillations $\hat H_1$ and temperature oscillations $\hat \theta_0$ across the heating cycle by taking the limit as $\tau_0 \rightarrow \infty$:

\begin{gather}
    \hat H_1 \equiv \lim_{\tau_0 \rightarrow \infty} H_1, \quad \hat \theta_0 \equiv \lim_{\tau_0 \rightarrow \infty} \theta_0.
\end{gather}

In this limit, the contributions from the initial conditions decay to zero and the oscillations repeat across the modulated time $\tau_0 \mod{\tau_{ch}}$. Next, we integrate Equation \ref{eq:Mixed Temperature Evolution} to introduce the boundary conditions into the long-term evolution. Finally, the sustained oscillations are incorporated by averaging across a single period $\tau_{ch}$. The resulting long-time surface and temperature evolution is given as:

\begin{gather}
    \pdev{H_0}{\tau_1} + \kappa \nabla_2 \cdot \left(\frac{1}{3} SH_0^3 \nabla_2(2 H_0 + \nabla_2^2 H_0) - \frac{1}{2} H_0^2 \nabla_2 \left[\tilde \theta_0 + \frac{V(\tau_{ch} - \tau_h - H_0^2)}{2H_0}\right]\right) = 0,  \label{eq:LT Surface Evolution}
    \\
    H_0\pdev{\tilde \theta_0}{\tau_1} -\frac{V(\tau_{ch} - \tau_h - H_0^2)}{2H_0} \pdev{H_0}{\tau_1} + \frac{\kappa}{\tau_{ch}}\int_{0}^{\tau_{ch}} \int_{-H_0}^{0}\left(U_{r0} \pdev{\hat \theta_0}{R} + U_{\gamma0} \pdev{}{\gamma}(\tilde \theta_0 + \hat \theta_0) \right) dR d\tau_0 \nonumber \\= -\left[\frac{1}{\tau_{ch}}\int_{0}^{\tau_{ch}}\hat H_1 \pdev{^2 \hat \theta_0}{R^2} d\tau_0\right]_{R=-H_0} 
    \nonumber \\
    - 4\left(\tilde \theta_0 + \frac{V(\tau_{ch} - \tau_h - H_0^2)}{2H_0} -\frac{a}{2}\int_0^{\gamma_0} \left(\tilde \theta_0 + \frac{V(\tau_{ch} - \tau_h - H_0^2)}{2H_0} \right) \sin \gamma \,d\gamma\right),
    \label{eq:LT Temperature Evolution}
\end{gather}

\begin{gather}
    \tilde U_r = \bm{\nabla}_2 \cdot \left(S \left(\frac{1}{6}R^3 + \frac{1}{2} H_0 R^2 \right) \bm{\nabla}_2 (2H_0 + \nabla_2^2 H_0) - \frac{1}{2}R^2 \bm{\nabla}_2 (\tilde \theta_0 + \hat \theta_0) \Big|_{R=-H_0} \right),
    \label{eq:LT Radial Velocity}
    \\
    \tilde U_\gamma = -S \left(\frac{1}{2}R^2 + H_0 R \right) \bm{\nabla}_2 (2H_0 + \bm{\nabla}_2^2 H_0 ) + R \bm{\nabla}_2 (\tilde \theta_0 + \hat \theta_0) \Big|_{R=-H_0},
    \label{eq:LT Polar Velocity}
\end{gather}
\begin{gather}
    \pdev{H_1}{\tau_0} = \kappa\nabla_2\cdot \left(\frac{1}{2} H_0^2 \nabla_2\left[\breve \theta_0 - \frac{V(\tau_{ch} - \tau_h - H_0^2)}{2H_0}\right]_{R=-H_0} \right). \label{eq:ST Surface Oscillation}
\end{gather}

\subsection{Linear Stability Analysis}

We derive the decay time of the long-term evolution by performing a linear stability analysis for small surface perturbations with amplitude $\eta$ away from a constant thickness film:

\begin{gather}
    H_0 = 1 + \eta H_{0,1} + O(\eta^2), \qquad H_1 = \eta H_{1,1} + O(\eta^2),  \\
    \tilde \theta_0 = \tilde \theta_{0,0} + \eta \tilde \theta_{0,1} + O(\eta^2).
\end{gather}

The leading order temperature parameter $\tilde \theta_{0,0}$ represents the long-time evolution of a constant thickness film : 

\begin{gather}
    \pdev{\tilde \theta_{0,0}}{\tau_1} = -4 F \left(\tilde \theta_{0,0} + K\right),
\end{gather}

where $F \equiv 1 - a (1 - \cos \gamma_0)/2$ is the proportion of the surface radiation emitted to the background, and $K \equiv V (\tau_{ch} - \tau_h - 1)/2$ is the average surface temperature of $\hat \theta_0$ for a constant thickness film ($H_0 = 1$). The solution to $\tilde \theta_{0,0}$ describes the radiative settling process, where the temperature oscillation will slowly change until the average surface temperature equals the reference temperature:

\begin{gather}
    \tilde \theta_{0,0} =  -K(1 - \exp(-4 F \tau_1)).
\end{gather}

At the next order, we evaluate the remaining integrals in Equations \ref{eq:LT Surface Evolution} and \ref{eq:LT Temperature Evolution} and express the results in terms of the harmonics of $H_{0,1}$ and $\tilde \theta_{0,1}$:

\begin{gather}
    \pdev{H_{0,1}}{\tau_1} + \frac{\kappa}{3}S\nabla_2^2(2H_{0,1}+\nabla_2^2 H_{0,1}) = \frac{\kappa}{2} \nabla_2^2\left(\tilde \theta_{0,1} - (K+V) H_{0,1} \right), \\
    \pdev{\tilde \theta_{0,1}}{\tau_1} + H_{0,1} \pdev{\tilde \theta_{0,0}}{\tau_1} - K \pdev{H_{0,1}}{\tau_1} + \kappa \left(C\nabla_2^2 H_{0,1} + \frac{1}{8} SV \nabla_2^2(2 H_{0,1} + \nabla_2^2 H_{0,1}) - \frac{1}{6} V\nabla_2^2 \tilde \theta_{0,1} \right) \nonumber \\= -\kappa J\nabla_2^2 H_{0,1} - 4 \left(\tilde \theta_{0,1} - (K+V) H_{0,1} - \frac{a}{2} \int_0^{\gamma_0} (\tilde \theta_{0,1} - (K+V) H_{0,1}) \sin \gamma d\gamma \right),
\end{gather}

where $C(\tau_{ch}, \tau_h)$ represents the impact of the heating cycle on thermocapillary convection, and $J(\tau_{ch}, \tau_h)$ represents the impact from small surface oscillations on the long-term evolution (see Supplementary Information for exact expressions of $C$ and $J$). The surface and temperature solution is then decomposed into spherical harmonic modes based on the Legendre function $P_l(x)$:

\begin{gather}
    H_{0,1} = A_l(\tau_1)P_l(\cos \gamma), \quad \theta_{0,1} = B_l(\tau_1)P_l(\cos \gamma) + W_l(\tau_1), \label{eq:Modes} 
\end{gather}

where $l$ is the eigenvalue of the mode, $A_l(\tau_1)$ and $B_l(\tau_1)$ are the amplitudes of the surface and temperature modes respectively, and $W_l(\tau_1)$ represents the uniform temperature offset produced by the self-radiation integral. As $W_l$ is independent of $\gamma$, it will not impact the surface evolution and thus can be isolated from the system:

\begin{gather}
    \frac{dW_l}{d\tau_1}= - 4 \left(F W_l - \frac{A_l - (K+V)B_l}{2} \int_0^{\gamma_0} P_l(\cos \gamma) \sin \gamma d\gamma \right).
\end{gather}

By using the Legendre function, the Laplacian reduces to a constant coefficient multiplying the mode: $\nabla_2^2 P_l(\cos \gamma) = -l(l+1) P_l (\cos \gamma)$. We then divide out all spatial components and express the long-term surface and temperature evolution in terms of the amplitudes $A_l$ and $B_l$:

\begin{gather}
     \frac{d}{d\tau_1}\sqmatrix{A_l \\ B_l} = (\Xi(\tau_1) +\Lambda_l)\sqmatrix{A_l \\ B_l},  \\
     \Xi(\tau_1) \equiv \sqmatrix{0 & 0 \\ 4FK\exp(-4F\tau_1) & 0}, \\
      \Lambda_l \equiv  -4\sqmatrix{0 & 0 \\ -K - V & 1} - \kappa l(l+1) \sqmatrix{
        -\frac{1}{2}(K+V) & \frac{1}{2} \\
        -\frac{1}{2} K (K+V)-C-J & \frac{1}{6}V + \frac{1}{2}K
    }\nonumber \\ - \frac{1}{3} S \kappa l(l+1)(l(l+1)-2) \sqmatrix{
        1 & 0 \\
        \frac{3}{8}V + K & 0
    }.
\end{gather}

\begin{figure}[hbtp]
    \centering
    \includegraphics[width=\linewidth]{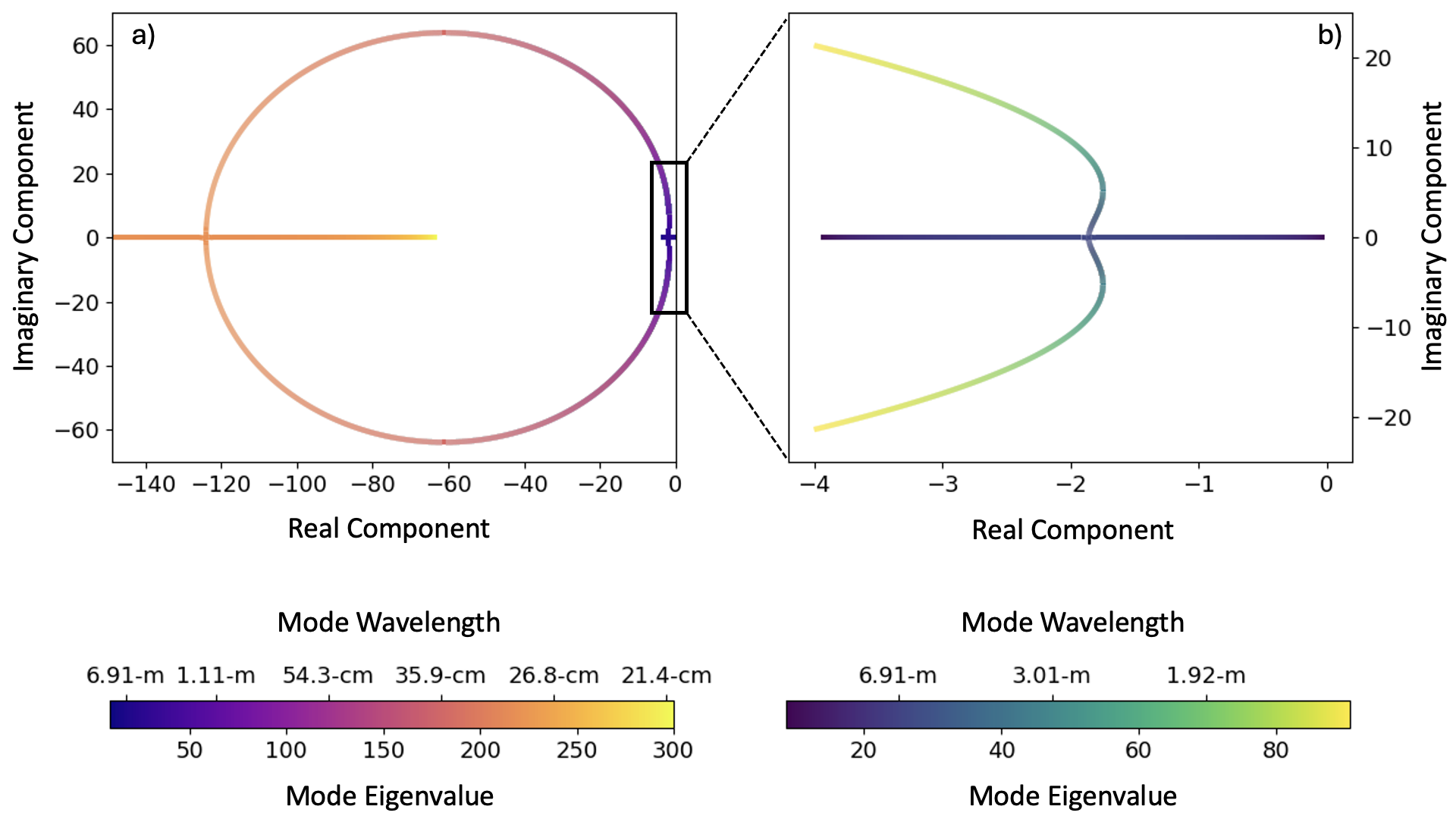}
    \caption{Root locus plot of the $\Lambda_l$ eigenvalues for different surface-temperature modes. Stable modes are indicated by negative real components, with larger values producing faster decay times. a) Eigenvalues across all three domains: the long-wave domain, mid-wave domain, and short-wave domain. Non-zero imaginary component results in oscillation. b) Eigenvalues of the long-wave domain and the beginning of the mid-wave domain. The small bump in the mid-wave domain arises from thermocapillary instability that is stabilized through radiation and capillary pressure.}
    \label{fig:RL Plot}
\end{figure}

The matrix $\Xi(\tau_1)$ represents the effect of radiative settling on the surface-temperature evolution, whereas the constant matrix $\Lambda_l$ describes how the surface and temperature distributions evolve through radiation, capillary pressure, and the thermocapillary effect. We observe that the evolution will be separated into two phases: the initial decay when $\Xi(\tau_1) \sim \Lambda_l$ during radiative settling, and the steady decay afterwards when $\Xi(\tau_1) \ll \Lambda_l$. Since $\Xi(\tau_1) \rightarrow 0$ as $\tau_1 \rightarrow \infty$, once radiative settling is complete, the stability and decay time of an individual mode depends on the eigenvalues of $\Lambda_l$, depicted in Figure \ref{fig:RL Plot}. The eigenvalues fall into three domains: the long-wave domain where the system is overdamped, the mid-wave domain where long-term thermocapillary oscillation begins, and the short-wave domain where capillary pressure dominates and overdampens the system. At the start of the mid-wave domain, we see a small increase in the real component of the eigenvalue and thus an increase in the stabilization time. This arises from thermocapillary instability emerging over the course of many cycles, becoming stronger at smaller wavelengths until it is quickly overtaken by capillary pressure.

\begin{figure}[hbtp]
    \centering
    \includegraphics[width=0.98\linewidth]{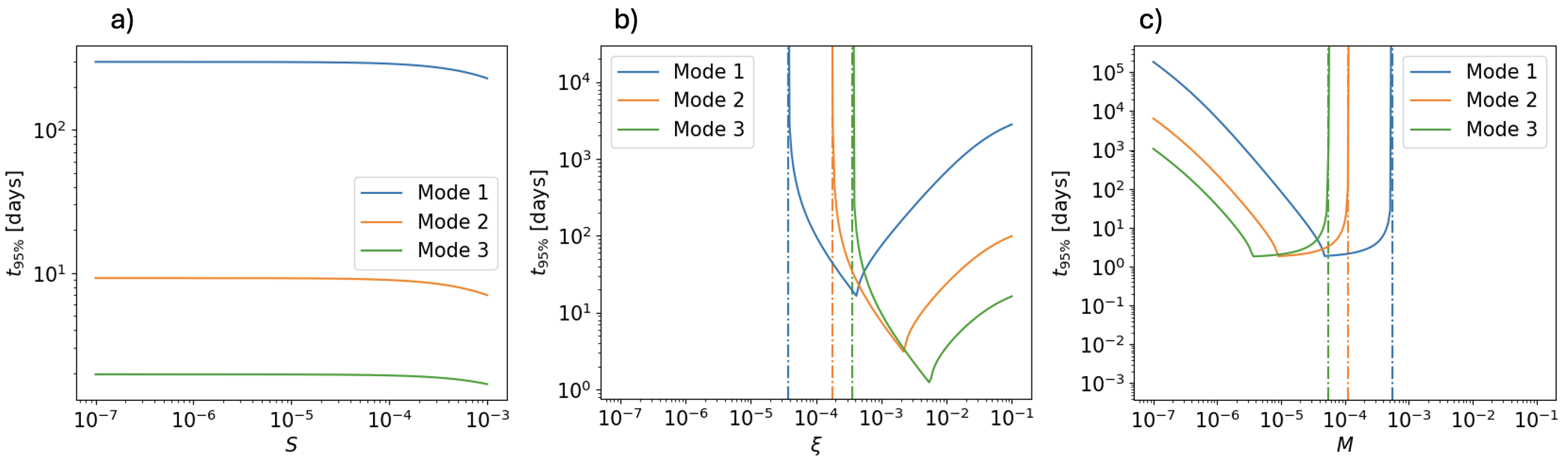}
    \caption{Stabilization time for first three modes relative to a) the inverse capillary number $S$, b) the radiative parameter $\xi$, and c) the scaled Marangoni number $M$. The vertical asymptote in b) and c) represent the onset of instability, and the cusps represent the transition between the long-wave and mid-wave domains.}
    \label{fig:System Parameter Variation}
\end{figure}

Here we define the 95\% stabilization time as $t_{95\%} \equiv -3h_{pin}^2 / \alpha\Re(\lambda_+)$, where $\lambda_+$ is the largest eigenvalue of $\Lambda_l \xi$. Figure \ref{fig:System Parameter Variation} shows a sweep of the system parameters $S$, $\xi$, and $M$ on the 95\% stabilization time of the first three modes of a 6.6-m liquid mirror. The scaled inverse capillary number $S$ has almost no impact on the long-wave modes, as capillary pressure is negligible at very long wavelengths.

We observe from the $\xi$ and $M$ sweeps in Figure \ref{fig:System Parameter Variation} that instability can occur in the mid-wave domain if $\xi$ is too small or $M$ is too large. When the system is underdamped, the stabilization time $t_{95\%, ud}$ is equal to:

\begin{gather}
    t_{95\%,ud} = \frac{3h_{pin}^2}{\alpha} \left(4\xi -  \frac{1}{3}M\left(V + 2 S \right)\cdot  l(l+1)  + \frac{1}{3} S M l^2(l+1)^2\right)^{-1}. \label{eq:Underdamped Decay Time}
\end{gather}

Instability will occur when $t_{95\%,ud} < 0$, resulting in the following stability criteria:

\begin{gather}
    M(V+2S)^2 < 48 S\xi .\label{eq:Stability Criteria}
\end{gather}

When this condition is violated, a range of unstable modes emerges between two critical eigenvalues:

\begin{gather}
    l_{c} = -\frac{1}{2} + \sqrt{\frac{1}{4} + \frac{M(V+2S) \pm \sqrt{M^2(V+2S)^2 - 48 SM\xi}}{2SM}}.
\end{gather}

From the $\xi$ and $M$ sweeps in Figure \ref{fig:System Parameter Variation}, we see a unique global minimum in stabilization time for each mode. The emitted radiation depends on the surface temperature, with hotter regions emitting slightly more energy than cooler regions. This results in radiation stabilizing the system by damping the transverse temperature gradients. The strength of this effect is determined by $\xi$, and reaches maximum stabilization when a mode is critically damped, represented by the global minimum. Modes can be targeted by turning the average surface temperature $T_{ref}$, changing the radiative parameter $\xi$ so that a given mode is critically damped and thus reaches its minimum stabilization time.

The parameter sweeps of $S$, $\xi$, and $M$ can be collapsed under the following transformation:

\begin{gather}
    S' \equiv S(l(l+1) - 2) \quad M' \equiv Ml(l+1)
    \\
    \Lambda_l\xi  \equiv  -4\xi \sqmatrix{0 & 0 \\ -K - V & 1} - M' \sqmatrix{
        -\frac{1}{2}(K+V) & \frac{1}{2} \\
        -\frac{1}{2} K (K+V)-C-J & \frac{1}{6}V + \frac{1}{2}K
    } \nonumber \\- \frac{1}{3} S'M' \sqmatrix{
        1 & 0 \\
        \frac{3}{8}V + K & 0
    }.
\end{gather}

\begin{figure}[hbtp]
    \centering
    \includegraphics[width=0.98\linewidth]{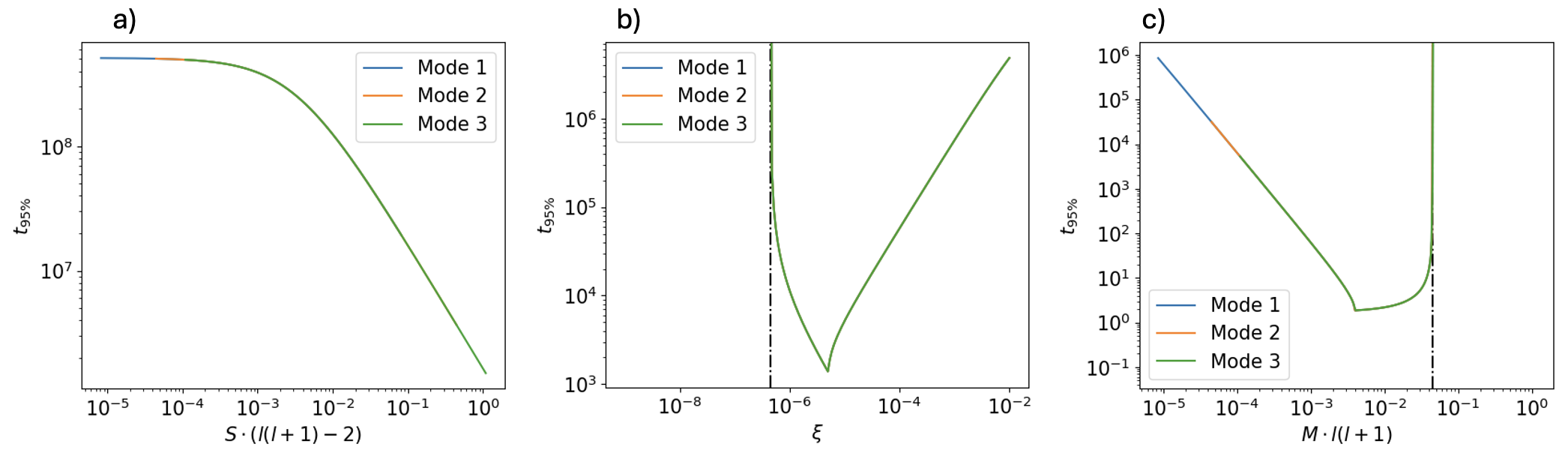}
    \caption{Collapsed stabilization curves for a) the transformed inverse capillary number  $S' = S(l(l+1) - 2)$, b) the radiative parameter $\xi$, c) the transformed Marangoni number $M' = M l(l+1)$.}
    \label{fig:Collapsed Curves}
\end{figure}

The collapsed curves from this transformation are shown in Figure \ref{fig:Collapsed Curves}. This transformation arises from the dependence on capillary pressure and mass transfer on the wavelength. The strength of capillary pressure depends on curvature from both variations in thickness as well as the underlying spherical curvature. The former scales with $l(l+1)$, whereas the latter does not depend on wavelength. These two factors combined give rise to the $l(l+1)-2$ scaling in $S$. As shorter wavelengths decrease transport distance and increase the strength of gradients, the rate of mass transfer $M$ scales with $l(l+1)$.

\begin{figure}[hbtp]
    \centering
    \includegraphics[width=0.98\linewidth]{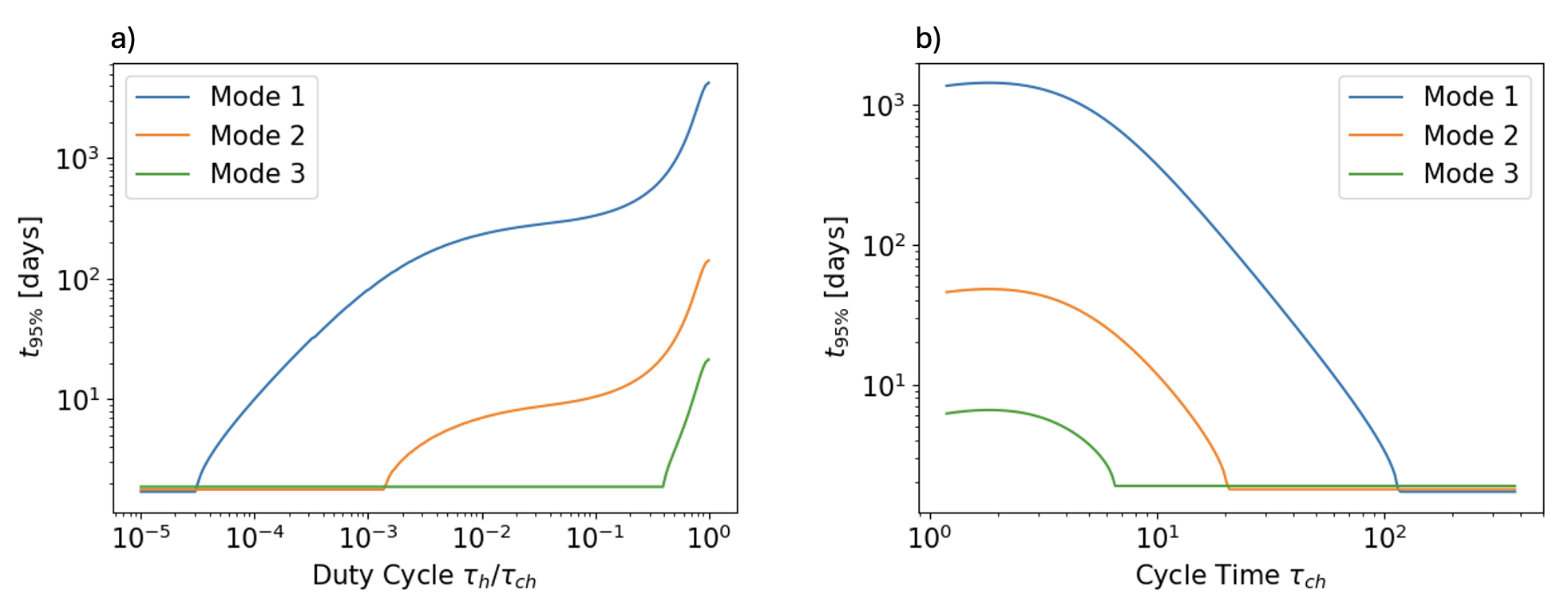}
    \caption{Stabilization time for first three modes relative to a) the duty cycle $\tau_h/\tau_{ch}$, b) the cycle time  $\tau_{ch}$ for a fixed duty cycle. Optimizing the heating cycle will improve long-wave modes up to the underdamped decay time, which is independent on the heating cycle. Variations in the final decay time between modes are due to convection and capillary pressure.}
    \label{fig:Heating Cycle Variation}
\end{figure}

Figure \ref{fig:Heating Cycle Variation} demonstrates that optimizing the heating cycle can only reduce the stabilization time of overdamped modes. While the heating cycle mitigates the formation of thermocapillary instability, the resulting oscillations will only decay through radiation, convection, and capillary pressure acting as damping mechanisms. For this reason, adjusting the heating cycle will only impact the long-wave modes up to the underdamped decay time in Equation \ref{eq:Underdamped Decay Time}.

\section{Numerical Simulations}

To validate the predicted stabilization times, we perform a numerical simulation of the long-wave model using the finite difference method and a backwards Euler scheme (see Supplementary Information for more information). To account for the moving boundary, the simulation is performed in a transformed coordinate system $(Z, G) \equiv (R/H, \gamma)$, fixing the boundaries to $Z \in [-1, 0]$ and $G \in [0, \gamma_0]$. The derivatives of $Z$ and $G$ are calculated using finite difference, then converted into the radial and polar derivatives:

\begin{gather}
    \pdev{}{R} = \frac{1}{H} \pdev{}{Z} \qquad \pdev{^2}{R^2} = \frac{1}{H^2}\pdev{^2}{Z^2} \\
    \pdev{}{\gamma} = \pdev{}{G} - \frac{Z}{H} \pdev{H}{G} \pdev{}{Z}\\
    \pdev{^2}{\gamma^2} = \pdev{^2}{G^2} - 2 \frac{Z}{H} \pdev{H}{G} \pdev{^2}{G\partial Z} + \left(\frac{Z}{H} \pdev{H}{G} \right)^2 \pdev{^2}{Z^2} + \left(\frac{2Z}{H^2} \left( \pdev{H}{G}\right)^2 - \frac{Z}{H} \pdev{^2H}{G^2} \right) \pdev{}{G}
\end{gather}

The temperature and surface systems are solved concurrently over each time. We focus our initial surface deformations to surface modes defined in Equation \ref{eq:Modes} whose amplitude starts at 1\% below the layer thickness with a constant starting temperature:

\begin{gather}
    H(\tau = 0) = 1 - 0.01 P_l(\cos \gamma) \qquad \theta(\tau=0) = 0
\end{gather}

\begin{figure}[hbtp]
    \centering
    \includegraphics[width=0.98\linewidth]{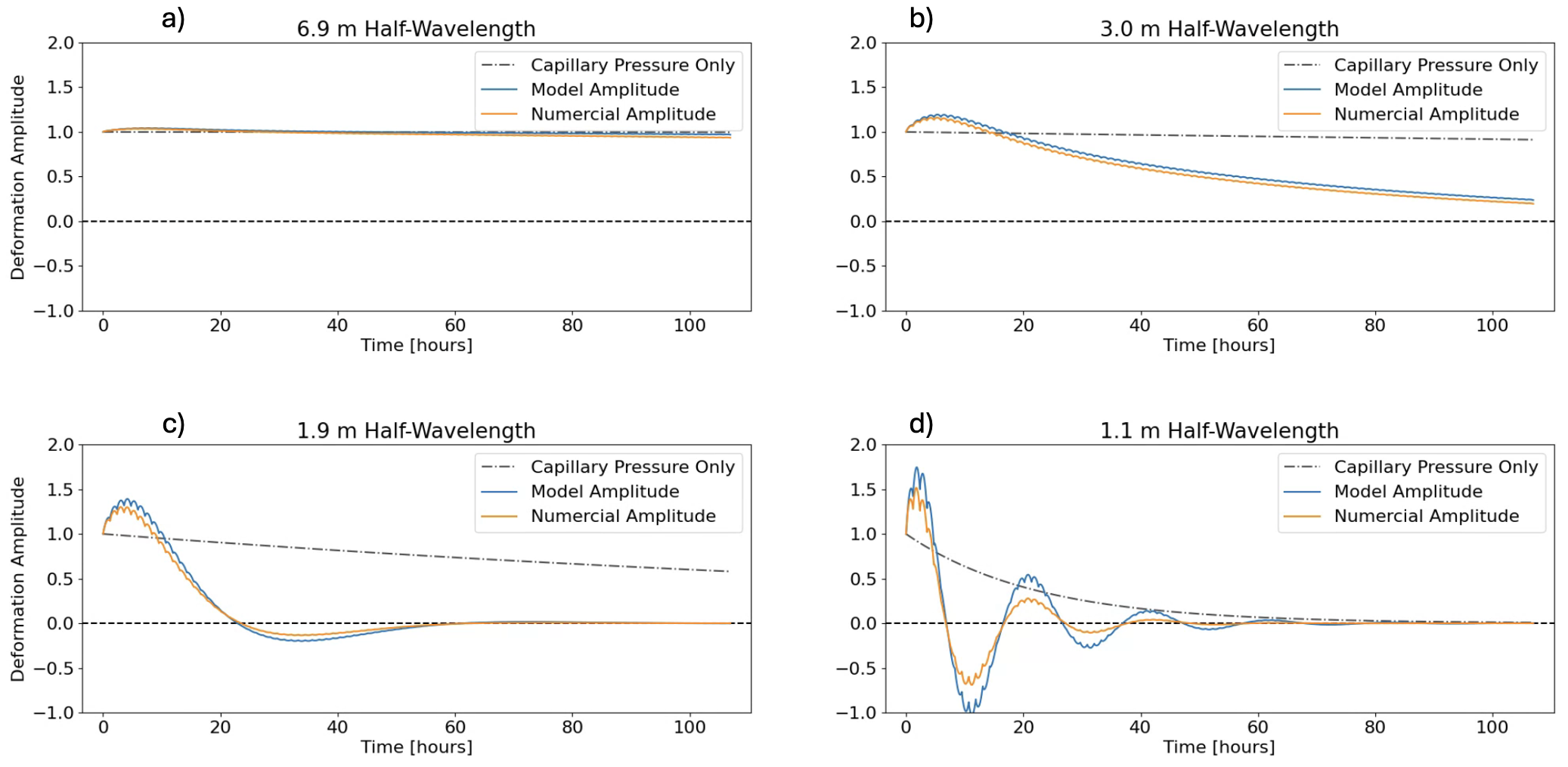}
    \caption{Amplitude evolution for modes a) 1, b) 2, c) 3, and d) 5. The model shows good agreement for the longer wavelengths, but overpredicts the stabilization time.}
    \label{fig:Numerical Results}
\end{figure}

The amplitude evolution of a 6.6-m diameter liquid mirror is shown in Figure \ref{fig:Numerical Results} for modes 1, 2, 3, and 5 representing the slowest decaying modes. All higher modes will decay significantly faster. We see good agreement between both the linear model and simulations across various parameters and modes tested. For a 5-K heating cycle, thermocapillary stabilization outpaces capillary pressure for all modes tested. Mode 2, corresponding to a 3.0-m half-wavelength at the center, demonstrates significant decay over 4.5 days with the perturbation predicted to reach 95\% decay after 8.6 days. In contrast, mode 1 (6.9-m half-wavelength) showed only an approximately 6\% decrease across the same time period with a predicted decay time of 9.1 months. To stabilize long-wavelength modes over the course of days instead of months, a larger temperature oscillation or a more complex heating cycle is required.

\section{Conclusion}

We have demonstrated thermocapillary stabilization of long-wave deformations for a spherical liquid layer in microgravity with surface radiation and modulated heating. The modulation prevents thermocapillary destabilization and allows radiation, convection, and capillary pressure to stabilize the free surface towards constant thickness. A basic heating cycle is able to stabilize large-scale deformations up to 3.0-m in half-wavelength over the course of days. The long-time evolution across many cycles was derived using multiple scale analysis,  and the decay times were then derived through a linear stability analysis. The resulting surface and temperature modes fall into three domains: the long-wave domain for overdamped modes, the mid-wave domain where modes exhibit long-time thermocapillary oscillations, and the short-wave domain where capillary pressure overdamps the system. The stabilization time is primarily determined by the strength of radiation and the rate of mass transfer through thermocapillary flow. If radiation is too weak or mass transfer is too quick, the system can become unstable based on the derived stability criteria. Heating cycle optimization can also improve the stabilization time for slow long-wave modes. Future work will investigate how introducing a more complex heating cycle can better target these long wave modes.

The multiple scale analysis presented here can be extended to other systems whose thermal boundary conditions weakly depend on temperature. The radiation boundary condition is the main damping mechanism, with hotter regions emitting more energy than cooler regions. Similar temperature-dependent boundary conditions such as convective cooling could provide additional stability to other heated thin-film systems. 

\section{Acknowledgments}
We thank Professor D. Holmes for the valuable discussions.

\subsection{Funding}
We acknowledge the financial support of the NASA Innovative Advanced Concepts (NIAC) program and NASA Ames Internal Research and Development (IRAD) fund.

\subsection{Data Availability}
The simulation source code, main script, and plotting script is available at \citep{citation-key}.

\subsection{Declaration of Interests}
The authors report no conflicts of interest.

\bibliographystyle{jfm}

\bibliography{FLUTE}

\end{document}